\documentclass[sigconf,screen]{acmart}

\AtBeginDocument{%
  \providecommand\BibTeX{{%
    \normalfont B\kern-0.5em{\scshape i\kern-0.25em b}\kern-0.8em\TeX}}}

\pdfoutput=1

\copyrightyear{2021} 
\acmYear{2021} 
\setcopyright{acmlicensed}\acmConference[ICAIF'21]{2nd ACM International Conference on AI in Finance}{November 3--5, 2021}{Virtual Event, USA}
\acmBooktitle{2nd ACM International Conference on AI in Finance (ICAIF'21), November 3--5, 2021, Virtual Event, USA}
\acmPrice{15.00}
\acmDOI{10.1145/3490354.3494372}
\acmISBN{978-1-4503-9148-1/21/11}

\usepackage{caption}
\usepackage{subcaption}
\usepackage{dsfont}
\usepackage{csquotes}
\usepackage{array}
\usepackage{soul}

\acmSubmissionID{33}

\begin{document} 
\title{Towards a fully RL-based Market Simulator}

\author{Leo Ardon}
\affiliation{%
  \institution{J.P. Morgan AI Research}
  \country{}
}
\email{leo.ardon@jpmorgan.com}

\author{Nelson Vadori}
\affiliation{%
  \institution{J.P. Morgan AI Research}
  \country{}
}
\email{nelson.n.vadori@jpmorgan.com}

\author{Thomas Spooner}
\affiliation{%
  \institution{J.P. Morgan AI Research}
  \country{}
}
\email{thomas.spooner@jpmorgan.com}

\author{Mengda Xu}
\affiliation{%
  \institution{J.P. Morgan AI Research}
  \country{}
}
\email{mengda.xu@jpmorgan.com}

\author{Jared Vann}
\affiliation{%
  \institution{J.P. Morgan AI Research}
  \country{}
}
\email{jared.vann@jpmorgan.com}

\author{Sumitra Ganesh}
\affiliation{%
  \institution{J.P. Morgan AI Research}
  \country{}
}
\email{sumitra.ganesh@jpmorgan.com}

\renewcommand{\shortauthors}{Ardon et al.}

\begin{abstract}
We present a new financial framework where two families of RL-based agents representing the Liquidity Providers and Liquidity Takers learn simultaneously to satisfy their objective. Thanks to a parametrized reward formulation and the use of Deep RL, each group learns a shared policy able to generalize and interpolate over a wide range of behaviors. This is a step towards a fully RL-based market simulator replicating complex market conditions particularly suited to study the dynamics of the financial market under various scenarios.
\end{abstract}



\maketitle

\section{Introduction}

Being able to understand the dynamics of the financial market has always been the Holy Grail of many economists and financial researchers. However, the number of actors at play and the variety of behaviors present in the market, make it very hard to fully understand the interplay between all the market participants. In the quest to solve the market making problem for instance, where the crux of the problem lies in understanding the interaction between the Liquidity Provider (LP), the Liquidity Taker (LT) and the Electronic Communication Network (ECN), two main approaches have emerged. The first one, more statistical, considers the different actors in isolation and make assumptions about the rest of the market. The other strategy is to model the market as a multi-agent system (MAS) where the participants are represented as independent entities able to interact between each other.

While this method seems more suited to help uncover phenomena emerging from the actions performed by multiple participants, modeling the agents' behavior is hard. Hand-coded policies driven by business experience and common sense are typically used but are either not complex enough to truly characterize the agent's behavior or are too difficult to calibrate with the data available.

With the recent breakthroughs in Reinforcement Learning (RL) and the development of RL-based multi agent systems, researchers have started to model the market participants as learning agents providing a less opinionated setting to study the dynamics of the market. By only specifying a well-designed reward function one can represent a wide range of behaviors without having to make hard assumptions on the policy to apply: the agent learns by itself what action to take in order to maximize its cumulative reward, based on the rules of the game only.

In this paper we build upon the work from \cite{ganesh2019reinforcement} with their use of RL to model the Liquidity Provider (or market maker) and introduce a new financial framework principally composed of RL-based agents able to learn how to react to a change in the market. We propose a unified reward formulation for the two groups of RL-based agents, each of which having its own family of behaviors balancing the trade-off between quantity and PnL. We leverage the theoretical framework from \cite{NEURIPS2020_a2f04745} to train concurrently two shared policies for the two families of agents present in the system, namely the Liquidity Takers and the Liquidity Providers. \emph{The parametrized formulation of the reward functions and the use of Deep Reinforcement Learning to train the shared policies, take full advantage of the generalization power of neural networks to learn over an entire spectrum of behaviors}.

\subsection{Related Work}

The work of Garman \cite{GARMAN1976257} was one of the first to present a stochastic model of the market making problem. He presented a framework where the market is centralized around one LP, who has the monopoly of all trading. He assumed that the aggregated supply and demand are exogenous to the LP and are Poisson distributed in time. This work was foundational to many subsequent research on the matter and paved the way to many discoveries. However, this work only considers a single LP and doesn't take into account the competition effect that multiple LPs have on each other. The fixed distribution of supply and demand does not truly reflect the reality of the financial market. In this configuration the LT are not able to react to the LP's actions. Building upon this theory, Amihud and Mendelson \cite{AMIHUD198031} and later on Guéant, Lehalle and Fernandez-Tapia \cite{gueant2013dealing} have focused on the importance of inventory (and the risk associated with it) to find the optimal pricing for market making. These papers tackle the problem of the temporal discrepancy between the orders executions and the continuous price change leading to the price risk a LP bears by holding inventory, but also ignore the competition effect among the agents.

The use of multi-agents systems to model the market participants has started to emerge more recently. In \cite{das2003agent}, Das successfully replicates important features of real financial time series by simulating the markets in a multi-agents system. The traders agents, corresponding to the LTs, are divided into two groups: the "informed" traders who have an idea of the fundamental value of the asset and trade accordingly; and the "liquidity" traders sometimes called Zero Intelligence (ZI) agents who trade randomly. These simple models of traders were sufficient to study simple financial facts but bring some limitations to the analysis of more complex market properties. The work of Cui and Brabazon (\cite{6327798}) and \cite{chan2001artificial} highlighted the necessity of intelligent agents to replicate real market conditions. In \cite{10.5555/2772879.2772890} and \cite{vyetrenko2019real} for instance, the authors successfully replicate some "stylized facts" observed in the markets by enhancing ZI agents with hand-coded heuristics. We argue that these heuristics only cover commonly known behaviors but do not capture more complex and undisclosed market strategies.

RL-based frameworks to solve the market making problem have been proposed. \cite{chan2001electronic} and more recently \cite{ijcai2020-633} and \cite{ganesh2019reinforcement} used reinforcement learning to train financial agent in a simulated market making environment. These approaches considered a single type of RL-based learning agent (the ECN or the LP) simplifying the model of the other market participants to heuristic based hand-coded policies. In our approach we relax the modeling assumptions made on the LT agent type to let emerge more natural behaviors.

\subsection{Our Contributions}

In this paper, \textbf{(i)} we formalize two types of market participants as RL-based agents, able to adopt different behaviors via a parametrized family of reward functions (sections \ref{sec:liquidity provider}, \ref{sec:liquidity taker}). \textbf{(ii)} We train simultaneously two groups of RL based financial agents, each using a shared policy (section \ref{sec:shared policy}) capable of interpolating over a range of behaviors. \textbf{(iii)} Finally, we perform an extensive study of the impact that the RL-based Liquidity Takers have on the pricing strategy of the Liquidity Providers (section \ref{sec:experiments}). In particular, we show that the Liquidity Provider agent learns to adapt his strategy depending on the type of investor he is connected to while the Liquidity Takers agent learns to react to different pricing strategy.

\section{Background} \label{sec:background}
We focus our research to the context of an order-driven market, where a single security is being traded for cash. Although many actors take part in this market, we can distinguish three types of participant: the Liquidity Provider (LP) also known as market maker, the Liquidity Taker (LT) or investor, and the Electronic Communication Network (ECN). The role of the Liquidity Provider is to ensure fluidity in the market by offering liquidity to buyers and sellers. LPs continuously quote \textit{bid} (buy) and \textit{ask} (sell) prices at which they are willing to trade. The LPs adapt their quotes dynamically to limit their exposure to the price variations caused by the fluctuation of supply and demand. Finding the optimal pricing strategy is refer to as \textit{market making}. The LP is connected to both LTs and ECNs and can trade directly with both of them. The LTs are the consumers; they enter the market to execute orders. They received the prices streamed by the LPs and ECNs and decide with whom they wish to trade. They typically trade directly with the LP because it offers better prices than going via the ECN.
The ECN plays an important role in the financial market by centralizing buy and sell orders. It can be seen as a collection of FIFO\footnote{First In First Out} queues of \textit{orders} placed at different levels by the market participants and waiting to be fulfilled. A level corresponds to the price at which the order will be executed. In general the ECN gives a good sense of how the market is evolving. It is often used as a proxy to represent all the other market participants.

\subsection{Notations}
\setlength{\arrayrulewidth}{0mm}
\renewcommand{\arraystretch}{1.5}
\begin{center}
\begin{tabular}{ | m{2.6cm} | m{4.9cm} | }
\hline
$\Delta x_{t} = x_{t} - x_{t-1}$ & The difference of the value of $x$ between two consecutive timesteps \\ 
\hline
$x_{t}^{(i)}$ & The value $x$ associated with the agent $i$ at the timestep $t$ \\ 
\hline
$\varepsilon = \frac{1}{2} \epsilon_{sym} + \epsilon_{asym}$ & The normalized tweak applied by the LP to the mid-price, relative to current ECN prices.
\end{tabular}
\end{center}

\subsection{Definitions}

\textbf{Prices}: The LP quotes bid and ask \textit{prices} at which it is willing to buy and sell, respectively. The ECN also exposes \textit{prices} on both the bid and ask sides. The best bid (respectively ask) corresponds to the highest (lowest) price at which a bid (ask) order can be executed on the ECN. We call \textit{mid-price} $P_{mid}$ the ECN mid-price; i.e the average of the best bid and best ask prices in the limit order book of the ECN. The mid-price is often use as a price of reference and the LP typically quotes their prices relative to $P_{mid}$.

\noindent\textbf{Market Spread}: The \textit{market spread} is the difference between the best ask price and the best bid price on the ECN LOB. We often use the half market spread $s^{ref}$ to reference the spread on each side. We can therefore write: 
\begin{align*}
    P_{ask}^{ref} &= P_{mid} + s_{ask}^{ref}; &
    P_{bid}^{ref} &= P_{mid} - s_{bid}^{ref};
\end{align*}

\noindent\textbf{Spread}: In the context of the LP, the \textit{spread} corresponds to the difference between the price quoted by the LP and the mid-price. In a similar fashion than above we can write the prices quoted by a LP $i$ as a function of the spread:
\begin{align}
    P_{ask}^{(i)} &= P_{mid} + s_{ask}^{(i)}; &
    P_{bid}^{(i)} &= P_{mid} - s_{bid}^{(i)}; \label{eq:spread}
\end{align}
We note however, that the spread $s_{ask/bid}^{(i)}$ is at the discretion of the LP, and different market making strategies will yield different spread. Depending of the quantity the LT wants to trade, the spread might differ.

\noindent\textbf{Inventory}: To accommodate the limited supply and demand, the LP typically maintains an inventory of the quantity traded until an investor accept to trade on the side at which it holds inventory. A positive inventory indicates that the LP has bought more than it has sold and vice versa. As presented in \cite{AMIHUD198031} and \cite{gueant2013dealing}, the \textit{inventory} plays an important role in the pricing of the LP.

\noindent\textbf{Risk aversion}: The risk aversion of an agent corresponds to a regularizer applied to the PnL, so as to penalize its fluctuations associated with carrying a large inventory. We choose to model the risk aversion as a penalty on the Inventory PnL, to capture the price risk the agents bear when holding inventory (\cite{gueant2013dealing}).

\noindent\textbf{Skewing}: With the aim of disposing of its inventory, the LP can make its prices on the side at which it holds inventory, more appealing to the LTs. The action of publishing asymmetric prices with one side more attractive than the other is called \textit{skewing}. The intensity of the skew is linked with the level of \textit{risk aversion} of the LP.

\noindent\textbf{Hedging}: Another way for the LP to reduce its inventory is to \textit{hedge}, i.e perform a trade which reduces its inventory. The LP takes the role of the LT in this configuration and therefore have to pay a cost.

\noindent\textbf{Spread PnL}: This excess in price between the bid and ask prices and the mid-price is paid by the LT. It can be seen as the cost the LT pays in order to trade the security. The profit or loss made by the LP by facilitating a trade is called the \textit{spread PnL}. It is a function of the quantity traded $q$ and the \textit{spread} at which the order was executed $s^{(i)}(q)$:
\begin{align*}
    \text{PnL}_{\text{spread}} = q \cdot s^{(i)}(q)
\end{align*}

\noindent\textbf{Inventory PnL}: As we have explained, holding inventory comes with profit or loss because of the frequent fluctuation of the mid-price. The \textit{inventory PnL} will be affected by the quantity of inventory $q$ and by the mid-price move $\Delta P_{mid}$:
\begin{align*}
    \text{PnL}_{\text{inventory}} = q \cdot \Delta P_{mid}
\end{align*}

\noindent\textbf{Total PnL}: The total PnL also simply referred as PnL, is the sum of \textit{Inventory PnL} and \textit{Spread PnL}.
\begin{align*}
    \text{PnL} = \text{PnL}_{\text{inventory}} + \text{PnL}_{\text{spread}}
\end{align*}

\section{Multi-Agent Framework} \label{sec:multi-agent framework}

\subsection{Configuration and Training}

To conduct our research we consider the dealer market as a \textit{Partially-Observable Stochastic Game} defined in \cite{DBLP:journals/corr/abs-2009-13051}; where each agent has only a partial view of the state of the world and aims at maximizing its own reward function by interacting with its environment. The different families of agents in our system are LP, LT and ECN. At each timestep, the LP updates the prices that it streams to the LT according to its observation of the market. Subsequently, and based on this new observation, the LT decides whether or not to trade with the LP or the ECN. The LP can also choose to hedge its position by trading with the ECN. The problem has thus been designed as a Stackelberg game, where the LT \textit{reacts} to the action that the LP takes (setting the prices it is willing to trade at).

As the focus of this paper lies in the interaction between the RL based LPs and LTs, we purposely give a brief description of the ECN model used in our experiments and leave a more extensive modeling for further research. Inspired by the work of Cont and Muller \cite{cont2021a}, we developed a statistical model to simulate the evolution of the order book as a function of its current state. The ECN agent is therefore able to react to the orders executed by the LPs and LTs and adjust its internal state to replicate the dynamics of the market. Our model is composed of three multivariate Gaussian mixtures calibrated using L2-level order book data (i.e snapshots of the ECN order book), and predicting the variation of the order book's snapshot over an interval of time $dt$. The first model is used to sample the initial snapshot of the book, that is, how much volumes is available at each level of the book. The second mixture models the variation of volume for each level as a function of the current volume. Finally, the third model decomposes the variation of volume to apply, into multiple smaller orders that will be added to the ECN.

\subsection{Shared Policy} \label{sec:shared policy}

The problem of multiple learning agents evolving in a decentralized partially-observable Markov decision process (DEC-POMDP) is known to be challenging \cite{10.5555/2073946.2073951}. Despite sharing the same environment, the agents have a different set of observations and act individually to maximize their own reward function. The already complex nature of the problem and the introduction of yet another set of learning agents to represent the Liquidity Takers can raise the question of the scalability of our approach and the training efficiency. We leverage the Centralized Training with Decentralized Execution technique presented in \cite{NEURIPS2020_a2f04745} to train two shared policies for the two families of agent present in our system. The use of a shared policy not only allow us to dramatically reduce the complexity of the problem by only having to train two neural networks; but also helps generalizing the trained policy by allowing the interpolation over the full behavior space of the agent family.

Within a family of agent $\Lambda$, each agent $i$ is of a specific type $\lambda^{(i)}$ characterizing the vector of parameters used by the reward function of the agent and defining its behavior. They all share the same action space $\mathcal{A}$ but in order to maximize their own cumulative reward $R_i$, they may take different actions $a_{t}^{(i)}$ due to their different states $s_{t}^{(i)}$ at any given time $t$ and their different reward function. The common policy $\pi_{\theta}^{\Lambda}$ shared by all the agents of the family $\Lambda$ is conditioned on the agent type $\lambda^{(i)}$ in order to allow generalization over the entire behavior space.

The policy is trained for a large number of episodes until convergence. The type $\lambda^{(i)}$ is sampled from a predefined distribution at the beginning of each episode and for each agent $i$. It is added to the observations of the agent as a constant value throughout the episode. The probabilistic nature of the agent type attribution enable the policy to learn across the spectrum of agent types.

\subsection{Liquidity Provider Agent} \label{sec:liquidity provider}

The role of the Liquidity Provider is to facilitate trading by providing liquidity to the other market participants. They continuously publish prices at which they are ready to buy and sell.

\subsubsection{Reward Function}\hspace{\fill}

We model the Liquidity Provider family as a spectrum of behaviors where on one end the agent tries to maximize its profit and on the other end it tries to gain a targeted share of the market. We propose a parametrized reward function able to define under a single formulation, a range of behaviors forming the Liquidity Provider family.

More rigorously the per timestep reward of a Liquidity Provider agent $i$ can be formulated as follows:
\begin{equation*}
    \text{Reward}_{t} = w \cdot \alpha \cdot \text{PnL}_{t}(\gamma) - (1-w) \cdot \text{MarketShare}_{t}(m^{*})
\end{equation*}

The parameter $w$ acts as a weight defining the importance allocated to the PnL or the the market share objective. An agent with $w=1$ will only trade with the aim to maximize its PnL. On the other hand, a weight $w=0$ will define an agent whose prime objective is to meet the market share objective irrespective of its PnL. The parameter $\alpha$ acts as a normalizer on the PnL component to make it comparable with the MarketShare objective.

The \textbf{PnL} component at an instant $t$ is defined by (\ref{eq:LP PnL}). The second part of the equation represents the risk aversion penalty associated with the inventory held. We call $\gamma$ the risk aversion of the LP.
\begin{equation}\label{eq:LP PnL}
    \text{PnL}_{t}^{(i)}(\gamma) = \Delta \text{PnL}_{t}^{(i)} - \gamma \big| \Delta \text{PnL}_{\text{inventory}, t}^{(i)} \big|
\end{equation}

The \textbf{Market Share} component of the reward function is defined in (\ref{eq:LP Market Share}). The goal is to minimize the difference between the empirical market share of the LP at the instant $t$ and the target $m^{*}$ specified as a parameter.
\begin{equation}\label{eq:LP Market Share}
    \text{MarketShare}_{t}^{(i)}(m^{*}) = \Delta \bigg| \frac{\sum_{s=0}^{t-1}m_{s}^{(i)}}{\sum_{j=0}^{N}\sum_{s=0}^{t-1}m_{s}^{(j)}} - m^{*} \bigg|
\end{equation}

This parametrized formulation provides enough flexibility to represent different behaviors. We can represent agents who want to increase their footprint in the market thanks to the \textit{Market Share Objective} component or the LPs who mainly trade to generate PnL with different level of risk aversion, or even a combination of both. By using the same reward formulation for all the LP agents and by incorporating the parameters of the reward function in the observation space, we train a \textit{shared policy} capable of interpolating over the behavior space of the Liquidity Provider.

\subsubsection{Pricing Formulation}\hspace{\fill}\label{sec:liquidity provider pricing formulation}

Market making is often seen as an optimal control problem (\cite{doi:10.1080/14697680701381228}, \cite{gueant2013dealing}) where the LP aims at finding the optimal prices to send to the LT in order to maximize its reward function. The LP quotes relative to the mid price and the problem then becomes finding the optimal spread to apply on the bid and ask sides. To model the spread applied by the LP agent, we decompose (\ref{eq:spread}):

\begin{align}
    P_{ask} &= P_{mid} + (s_{ask}^{ref}+s_{bid}^{ref}) \bigg[\frac{1}{2} (1+\epsilon_{sym}) + \epsilon_{asym} \bigg] \label{eq:additive spread ask} \\
    P_{bid} &= P_{mid} - (s_{ask}^{ref}+s_{bid}^{ref}) \bigg[\frac{1}{2} (1+\epsilon_{sym}) - \epsilon_{asym} \bigg] \label{eq:additive spread bid}
\end{align}

The two newly introduced parameters $\epsilon_{sym}$ and $\epsilon_{asym}$ will be learnt by the policy network as a function of the observations available to the RL agent. As we can see in (\ref{eq:additive spread ask}) and (\ref{eq:additive spread bid}), $\epsilon_{sym}$ is a \textit{symmetric} tweak of the prices around the mid-price, it can be thought of as the parameter controlling the willingness of the LP to win the flow. A negative $\epsilon_{sym}$ indicates that the LP offers to trade at a price more favorable for the LT, generating less Spread PnL for the LP but with a higher probability of being executed. $\epsilon_{asym}$ is an \textit{asymmetric} tweak and is used for skewing (i.e making one side more attractive than the other to investors) with the objective of reducing the inventory held.

\subsubsection{Agent Configuration}\hspace{\fill}

\textbf{Actions:} At each timestep, the RL agent characterizing a LP can decide the value for three parameters:
\begin{itemize}
    \item[$\blacksquare$] Symmetric price tweak $\epsilon_{sym}$
    \item[$\blacksquare$] Asymmetric price tweak $\epsilon_{asym}$
    \item[$\blacksquare$] Fraction $h$ of its inventory to hedge
\end{itemize}

\textbf{Observations:} For the agent to learn the optimal policy the following observations are passed to the policy network at each timestep:
\begin{itemize}
    \item Reference mid-price $P_{mid}$ and its history 
    \item Inventory $z$ currently held
    \item Fraction of time elapsed since the start of the simulation
    \item Empirical Market Share $m$ the LP had in the previous timestep
    \item The top $n$ levels of the ECN order book
    \item Cost associated with hedging for multiple hedge fractions $h$
\end{itemize}

The parameters characterizing the agent type are also added to the observations:
\begin{itemize}
    \item Weight $w$ associated with the PnL component
    \item Risk aversion parameter $\gamma$
    \item Market Share target $m^{*}$
    \item Empirical probabilities of being connected with a LT and a ECN agent
\end{itemize}

\subsection{Liquidity Taker Agent} \label{sec:liquidity taker}

The Liquidity Taker, also called investor is an actor who enters the market to trade. As opposed to the Liquidity Provider, it doesn't publish quotes; rather it consumes the prices streamed by the LP and makes the choice on whether or not they wish to trade with it, but do not have any obligation to do so. We innovate by introducing a RL-based agent to take the role of the Liquidity Taker in our market simulator. We start by formalizing the Liquidity Taker reward function and present a toy example demonstrating the different behaviors a LT can adopt.

\subsubsection{Reward Function}\hspace{\fill}

Similarly to the LP, the behavior a LT adopts can be represented as a combination of two components: a PnL component and a Target Flow component. The PnL part is analogous to the one presented for the LP, the agent aims at maximizing its PnL and execute trade optimally for that purpose (Informed traders are a typical example of PnL driven LT where the PnL component predominates). On the other end of the spectrum, some LT are more flow driven. Their primary objective is to trade a targeted quantity. The target Flow indicates how much they should trade on each side. To reduce the complexity of the LT agent, we assume that it always trades a fixed unit quantity. This simplifies the Flow objective component to only track the frequency for each action. The formulation also supports mixed behavior where the agent tries to satisfy its trade objective with a certain tolerance, allowing it to generate more PnL.

The formulation below defines a family of agents where a given agent $i$ could fall anywhere on that spectrum. The parameter $w$ is used as weight to make the agent more "PnL" or "Flow" driven.
\begin{equation*}
    \text{Reward}_t = w \cdot \alpha \cdot \text{PnL}_{t}(\gamma) - (1-w) \cdot \text{Flow}_{t}(\mathbf{q}^{*})
\end{equation*}

The \textbf{PnL} component is the same as the one presented in (\ref{eq:LP PnL}) where a penalty is added to the PnL in order to accommodate for the risk associated with holding inventory:
\begin{equation}
\text{PnL}_{t}^{(i)}(\gamma) = \Delta \text{PnL}_{t}^{(i)} - \gamma \big|\Delta \text{PnL}_{\text{inventory}, t}^{(i)} \big| \tag{\ref{eq:LP PnL}}
\end{equation}

The \textbf{Flow} objective of the reward function evaluates the distance between the flow executed by the agent up until $t$ and the flow objective $\mathbf{q}^{*}$ specified as parameter. In practice, the objective is an array containing the targets for each of the the possible actions $a$ of the action space $\mathcal{A}$.

\begin{equation}\label{eq:LT Flow}
\begin{gathered}
\text{Flow}_{t}^{(i)}(\mathbf{q}^{*}) = \frac{1}{|\mathcal{A}|} \sum_{j \in \mathcal{A}} \Delta \bigg| \frac{1}{t}\sum_{s=0}^{t-1}\mathds{1}_{\{a_t^{(i)} = j\}} - q_{j}^{*} \bigg| \\ 
\sum_{j \in \mathcal{A}} q_{j}^{*} = 1
\end{gathered}
\end{equation}

\begin{figure}[ht]
    \centering
    \begin{subfigure}[b]{0.23\textwidth}
        \centering
        \includegraphics[width=\textwidth]{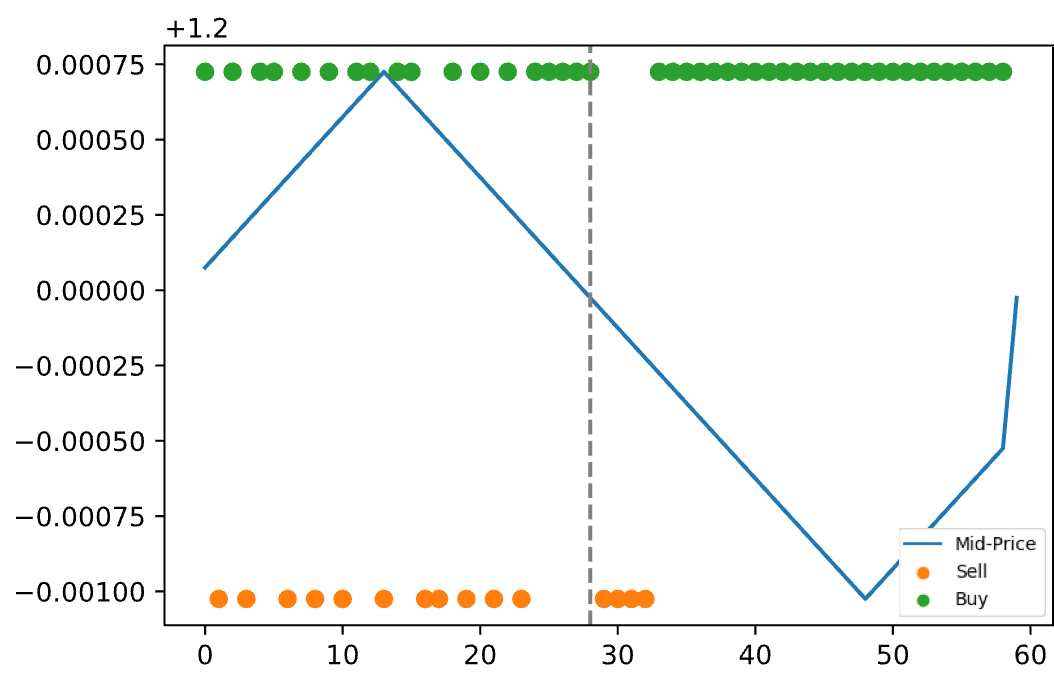}
        \caption{$w=0.00$}
        \label{fig:w=0.0}
    \end{subfigure}
    \hfill
    \begin{subfigure}[b]{0.23\textwidth}
        \centering
        \includegraphics[width=\textwidth]{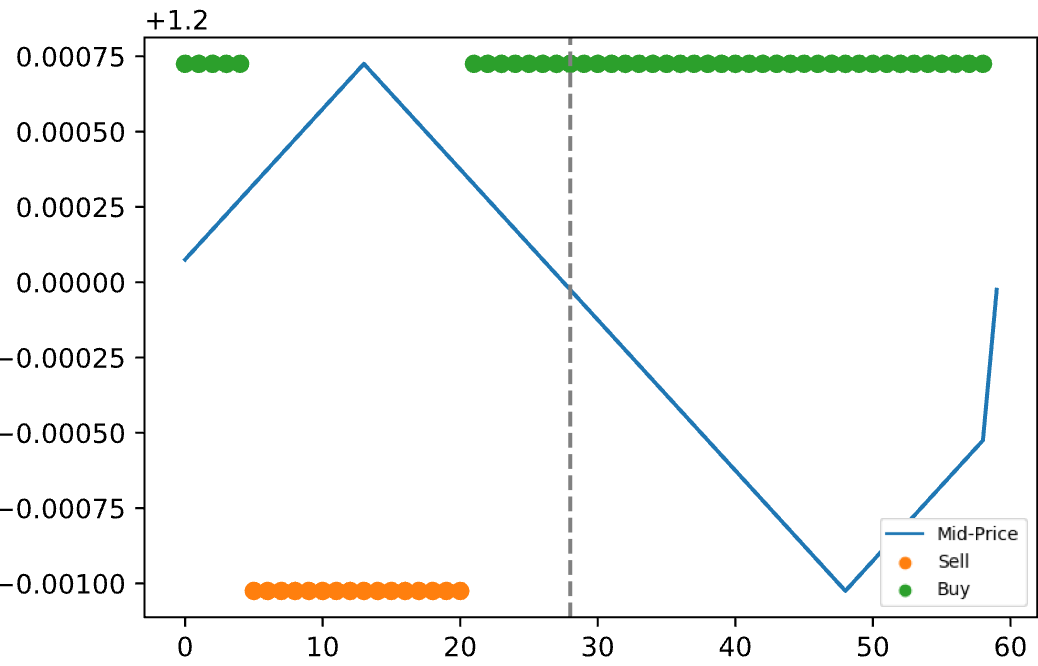}
        \caption{$w=0.25$}
        \label{fig:w=0.25}
    \end{subfigure}
    \hfill
    \begin{subfigure}[b]{0.23\textwidth}
        \centering
        \includegraphics[width=\textwidth]{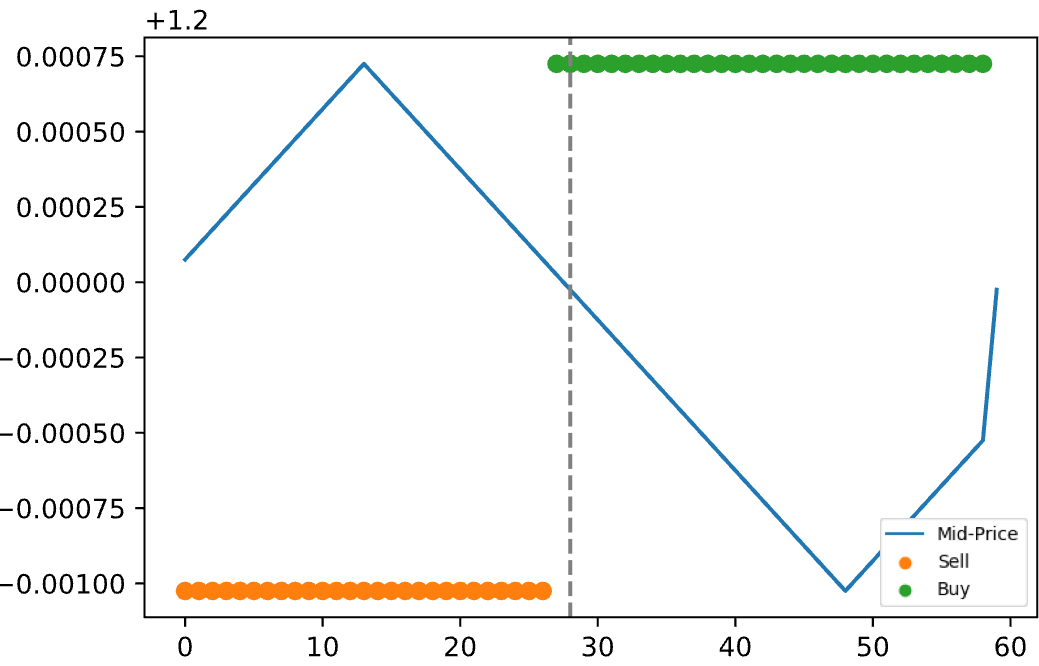}
        \caption{$w=0.75$}
        \label{fig:w=0.75}
    \end{subfigure}
    \hfill
    \begin{subfigure}[b]{0.23\textwidth}
        \centering
        \includegraphics[width=\textwidth]{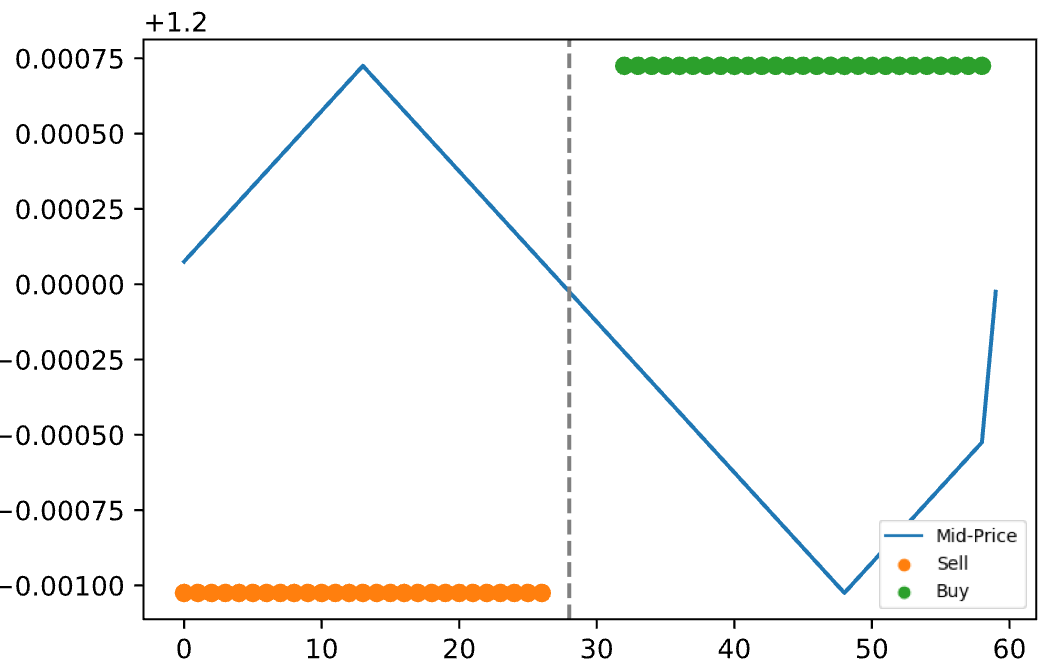}
        \caption{$w=1.0$}
        \label{fig:w=1.0}
    \end{subfigure}
    \caption{Spectrum of behavior for the Liquidity Taker}
    \label{fig:LT toy example}
    \small{
    With this simple example where we have artificially added a trend in the mid-price, we see the different behaviors the LT can adopt depending on its reward function. As the weight $w$ associated with the PnL component increases, the LT progressively trades more and more optimally to maximise its PnL at the expense of its target flow of 75\% Buy-25\% Sell.}
\end{figure}

To illustrate the variety of behaviors of this type of agent, we present in \textit{Figure \ref{fig:LT toy example}} a toy example where an artificial trend is added to the mid-price. Adding this trend allows us to easily identify the best trading strategy the agent should learn in order to maximize its profit. The idea is to observe how the behavior of the LT agent evolves as the weight $w$ applied to the PnL component changes. For this example we use a simple configuration of our simulator with one LP and one LT. We parametrize the LT agent with no risk aversion $\gamma = 0.0;$ and with an objective to sell 25\% of the time and buy 75\% of the time: $q_{ask}^{*} = 0.25; \; q_{bid}^{*} = 0.75$.

As we can observe in \textit{Figure \ref{fig:w=0.0}} when the principal objective for the LT is to meet the specified flow objective, the agent trades with no particular consideration of the price. It successfully buys 75\% of the time and sells 25\%. With the weight on the PnL component increasing the agent learns to adapt its trading strategy to buy and sell when it is the most favorable to generate PnL. For $w=1$ in \textit{Figure \ref{fig:w=1.0}}, the LT learnt to trade perfectly following the famous motto \textquote{buy low, sell high}. Intermediate behaviors are shown in \textit{Figure \ref{fig:w=0.25},  \ref{fig:w=0.75}} where we see that the agent gradually discard its objective of matching its flow targets and focus more on maximizing its PnL.

\subsubsection{Agent Configuration}\hspace{\fill}

\textbf{Actions:} At every timestep, the RL-based Liquidity Taker agent can choose to \textit{buy}, \textit{sell} or \textit{not trade} in order to maximize its reward function.

\textbf{Observations:} At each timestep, the policy network of the LT agent is fed with the observations below:
\begin{itemize}
    \item Reference mid-price $P_{mid}$ and its history 
    \item Inventory $z$ currently held
    \item Fraction of time elapsed since the start of the simulation
    \item Proportion of the flow the LT has traded on each side $q_{ask}$ and $q_{bid}$
    \item Cost associated with executing a trade on both side for each LP the LT is connected to
\end{itemize}

We also enrich the observations with the parameters of the reward function and the connectivity of the agent type for the shared policy to learn the distribution of behaviors characterizing the LT:
\begin{itemize}
    \item Weight $w$ associated with the PnL component
    \item Risk aversion parameter $\gamma$
    \item Flow targets $q_{ask}^{*}$ and $q_{bid}^{*}$
    \item Empirical probabilities of being connected with a LP and a ECN agent
\end{itemize}

\section{Experiments} \label{sec:experiments}

We focus our attention on the empirical study of the effect that the different behaviors of LT can have on the LP's actions. We aim at answering the following questions with our study: 1) Can the two groups of agents learn simultaneously? 2) How does the LP agent adapt its behavior in the presence of different types of LT? 3) Do we need this new type of agents to replicate observed market properties?

We present below the results from various experiments trying to approach the problem from different angles. In the first configuration of experiments, we change the proportion of LT actors of a certain type in the system. We start with a similar configuration than the one presented in \cite{ganesh2019reinforcement} and gradually increase the number of PnL driven investors. We try to evaluate how the heterogeneity of behaviors can affect the LP's actions. The second configuration of experiments takes the approach of varying the connectivity between the agents of a particular type. The study of the connectivity is important because two agents not connected have still an indirect effect on each other by interacting with and modifying the environment.

Using the RLlib library \cite{pmlr-v80-liang18b} and the OpenAI Gym framework \cite{brockman2016openai} we developed a multi-agent market environment, allowing us to put in competition different types of RL based market participants. The shared policies of the agents are trained using the standard RLLib implementation of the Proximal Policy Optimization (PPO) algorithm which supports parallel execution providing us with the scalability needed for a complex environment like ours.

\subsection{The Impact of Diversity}

\subsubsection{Price Tweak Distribution} \label{sec:price tweak distribution}\hspace{\fill}

\begin{figure}[ht]
    \centering
    \includegraphics[width=0.47\textwidth]{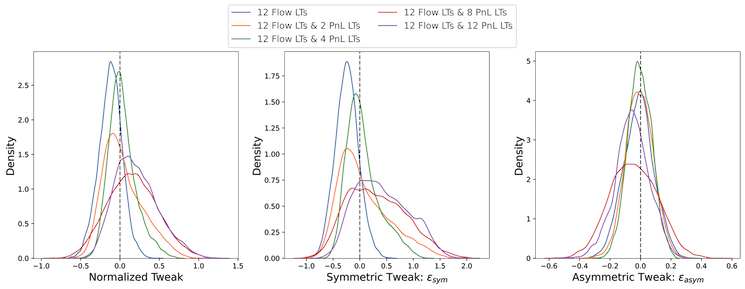}
    \caption{Distribution of the price tweak}
    \label{fig:Skew Distribution}
\end{figure}

\noindent \textbf{Setup:} We present in \textit{Figure \ref{fig:Skew Distribution}} the distribution of the normalized tweak $\varepsilon$ applied by the LP to its prices in order to attract LT. As presented in \ref{sec:liquidity provider pricing formulation} (\ref{eq:additive spread ask}) and (\ref{eq:additive spread bid}), the spread applied to the ECN mid price have a symmetric $\epsilon_{sym}$ and an asymmetric tweak $\epsilon_{asym}$, we therefore also plot the distribution for these two components. We ran multiple experiments varying the proportion of Flow and PnL driven LTs in the system. The experiments are run with 3 LPs, 12 Flow driven LTs and an increasing number of PnL driven LTs (0, 2, 4, 8, 12). The goal of these experiments was to understand the effect of the proportion of different LT types on the way the LPs price. A \textit{negative} normalized tweak $\varepsilon$, indicates that the LP is offering better prices than the ECN and the more negative the tweak is, the more attractive the prices will be to investors. For the asymmetric component of the tweak $\epsilon_{asym}$, a \textit{negative} value indicates that the \textit{ask} side is favoured and a \textit{positive} value favours the \textit{bid} side. 

\noindent \textbf{Analysis:} As the proportion of PnL driven LTs increases, we see a shift in the distribution of the normalized tweak $\varepsilon$ towards higher values indicating that the LP published worse prices. Looking at the two components of the tweak, we observe that this shift comes from the symmetric component $\epsilon_{sym}$. The LP's PnL suffers when too many informed LTs are in the system. To compensate its losses, \textbf{the LP becomes more conservative in its pricing as the proportion of PnL driven LTs increases}. The asymmetric tweak $\epsilon_{asym}$ is centered around the origin implying that, as expected, the proportion of PnL driven LTs has no impact on the direction of the skew. However, we do observe fatter tails for this distribution suggesting an increased skewing intensity from the LPs on one side or the other, which can also be interpreted as a \textbf{higher risk aversion}. Since both the LPs and the PnL driven LTs try to maximize their Inventory PnL, as the proportion of PnL LTs increases the adversarial effect on the LPs intensifies and their Inventory PnL get lower. 

\subsubsection{Flow by LT Agent type}\hspace{\fill}

\begin{figure}[ht]
    \centering
    \includegraphics[width=0.43\textwidth]{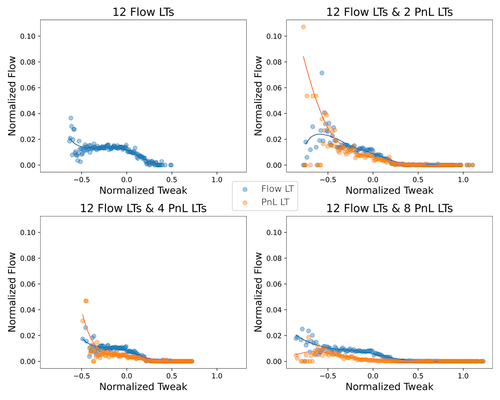}
    \caption{Flow by LT type}
    \label{fig:Skew vs Flow}
\end{figure}

\noindent \textbf{Setup:} In \textit{Figure \ref{fig:Skew vs Flow}} we plot the average flow between a LP and the LT of different types (Note that both the bid and ask sides are combined on the same chart). The reward function of the PnL driven LT being different than the reward function of the Flow driven LT, they should adopt different behaviors and have different reactions to the way the LP tweaks its prices. We expect the PnL driven LT to be more demanding and only trade for good prices. On the other hand, because their goal is to meet a flow objective, the Flow driven LT needs to trade with little respect to the price. For better prices we should therefore see a higher flow coming from the PnL driven LTs. We run multiple experiments with different proportions of PnL and Flow LTs.

\noindent \textbf{Analysis:} As we can see in the top left quadrant of \textit{Figure \ref{fig:Skew vs Flow}}, a configuration with only Flow driven LT does not help replicating the "stylized facts" observed in the market. The flow between the LP and the LT increases slowly as the prices get better for $\varepsilon \in [0.25, 0.00]$. With $\varepsilon \geq 0.25$, the prices streamed by the LP are so bad that it doesn't attract any flow and the LT prefer to trade with the ECN instead as it offers better prices. We observe that with $\varepsilon \leq 0.00$, the flow plateaus with a slight uptake for very good prices. With the introduction of PnL driven LTs, we can see in the top right quadrant, what we were expecting: because they only trade when the price is best for their PnL, \textbf{the PnL driven LTs see an exponential uptake of their flow as the prices improve}. However, as we can observe in the bottom charts, as we increase the number of PnL driven LTs the convexity in the flow uptake shrinks and with too many informed LTs, their flow becomes almost null. This is associated with the results presented in \ref{sec:price tweak distribution} where too many PnL LTs make the LP tweak its prices less, not making it worthwhile for the PnL LT to trade.

\subsubsection{PnL by Agent type}\hspace{\fill}

\begin{figure}[ht]
    \centering
    \includegraphics[width=0.42\textwidth]{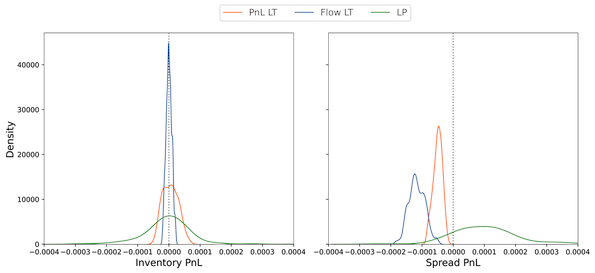}
    \caption{Spread and Inventory PnL by Agent type}
    \label{fig:PnL Distribution}
\end{figure}

\noindent \textbf{Setup:} For the configuration of 12 Flow and 2 PnL driven LTs, we now try to gain more insights about the different types of agents. We present in \textit{Figure \ref{fig:PnL Distribution}} the average Spread and Inventory PnL for each type. Looking at the PnL helps us understand their trading pattern. We wish to understand how each agent type performs against the others. As we have explained previously the Flow and PnL LTs use a shared policy learning a family of reward functions. The LPs use a shared policy of their own, which means that we are in a situation where two shared policy are competing. We recall that when the LT trades with a LP, it pays the \textit{spread}, generating a positive Spread PnL for the LP but negative for the LT. In these experiments the flow objective of the Flow driven LT is to trade 50\% of the time on the bid side and 50\% on the ask side.

\noindent \textbf{Analysis:} Looking at the Inventory PnL in \textit{Figure \ref{fig:PnL Distribution}}, we see that the distribution of PnL for all 3 agent types is centered around the origin. The distribution for the Flow LT is narrower than the others because of their objective to trade frequently. Consequently they do not hold a large inventory and therefore generate less Inventory PnL. As expected the PnL driven LT has a wider distribution of Inventory PnL indicating that it earns more PnL by strategically holding inventory. However, the fact that it is only connected to the ECN and the LPs, doesn't allow the PnL LT to generate as much Inventory PnL than the LP. The LP benefits from its connectivity with the Flow LTs providing the flow required to build inventory. The analysis of the Spread PnL distribution also offers interesting insights. As we were expecting, because the LT pays the spread, its Spread PnL is negative. The Spread PnL of the LP is mostly positive but is sometimes negative as the LP pays the spread when it hedges itself at the ECN. The Spread PnL of the PnL driven LT is significantly higher than the Spread PnL of the Flow LT. The PnL LT agent learns to trade with the objective of maximizing its PnL, it therefore tries to trade only when the Spread is more favourable, generating a higher spread PnL.

\subsection{Role of connectivity with Flow LTs}

In the next set of results presented below, we take a different approach and play with the connectivity between the Flow LTs and the LP. Since an agent can only trade with a connected agent, its connectivity with the other families implicitly have an effect on the agent's behavior and thus is part of its type. Because of their nature, the Flow driven LT, have to trade to meet their flow objective irrespective of the price. That makes them a guaranteed source of flow for the LP and therefore a reliable source of Spread PnL. By varying the connectivity between these two types of agent and in the presence of PnL driven LTs, we wish to understand how the behavior of the LP is affected.

\subsubsection{Pricing as a function of connectivity}\hspace{\fill}

\begin{figure}[ht]
    \centering
    \includegraphics[width=0.32\textwidth]{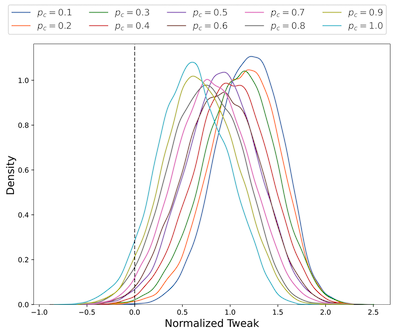}
    \caption{Distribution of the price tweak for different level of connectivity between Flow LTs and LPs}
    \label{fig:Skew connectivity}
\end{figure}

\noindent \textbf{Setup:} In \textit{Figure \ref{fig:Skew connectivity}}, we display the distribution of the tweak applied by the LP in its prices as a function of the empirical probability of being connected to a Flow LT. As before, the lower the price tweak, the more attractive the prices will be for LTs. The goal of this experiment is to understand how the LP adapts its behavior when its guaranteed source of flow is limited because of a lower probability of being connected.

\noindent \textbf{Analysis:} A clear trend appears on \textit{Figure \ref{fig:Skew connectivity}}, as the probability of being connected decreases the LP becomes more conservative in its prices. In fact, with low connectivity with Flow driven LTs, it gets more and more difficult to dispose of its inventory. Consequently, to accommodate for the risk associated with holding inventory the LP publishes worse prices. We observe a similar effect than in \ref{sec:price tweak distribution}, indicating that reducing the connectivity between LP and Flow LTs yield the same behavior than having a higher proportion of PnL driven LTs.

\subsubsection{Skew intensity}\hspace{\fill}

\begin{figure}[ht]
    \centering
    \includegraphics[width=0.28\textwidth]{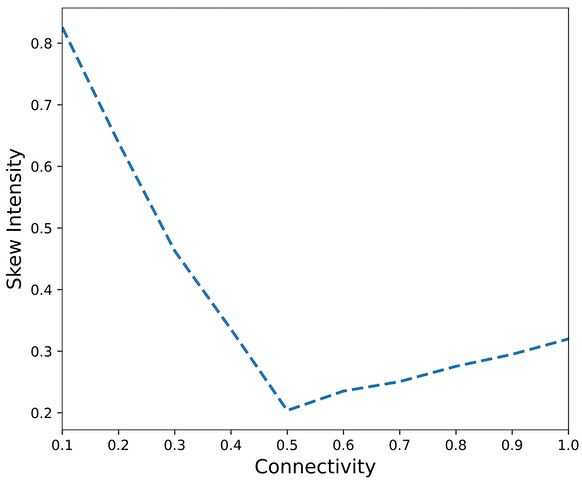}
    \caption{Skew intensity as a function of Connectivity}
    \label{fig:Skew Intensity}
\end{figure}

\noindent \textbf{Setup:} We now study the effect of connectivity on the intensity of the skew the LP applies to its prices, that is how much the LP tweak its prices asymmetrically to attract the investors on the side at which it holds inventory in order to reduce its risk. We show on \textit{Figure \ref{fig:Skew Intensity}}, the skew intensity as a function of the connectivity. With a lower probability of being connected with a Flow LT, it will be more challenging for the LP to manage its inventory because of the absence of flow guarantee the Flow LT provides. The LP is thus exposed to an unfavorable price move that would impact its Inventory PnL. We therefore expect the LP to be more willing to get rid of its inventory for lower connectivity which would be materialized by a higher skew in its prices.

\noindent \textbf{Analysis:} As we can see in \textit{Figure \ref{fig:Skew Intensity}}, the probability of being connected to a Flow LT affects the way the LP prices. \textbf{With low connectivity, the LP is more aggressive in its skew}. The incertitude about the ability to dispose of its inventory is forcing the LP to skew more its prices to increase the chances of attracting investors. As the connectivity increases, the LP reduces the intensity of the skew linearly up until a probability of $0.5$ of being connected, after which the intensity remains relatively stable until full connectivity.

\subsection{Risk aversion}

In this last experiment, we evaluate the behavior of the LP with different level of risk aversion. We keep the configuration of LTs constant and only change the risk aversion factor $\gamma$ of the reward function of the LP.

\begin{figure}[ht]
    \centering
    \includegraphics[width=0.34\textwidth]{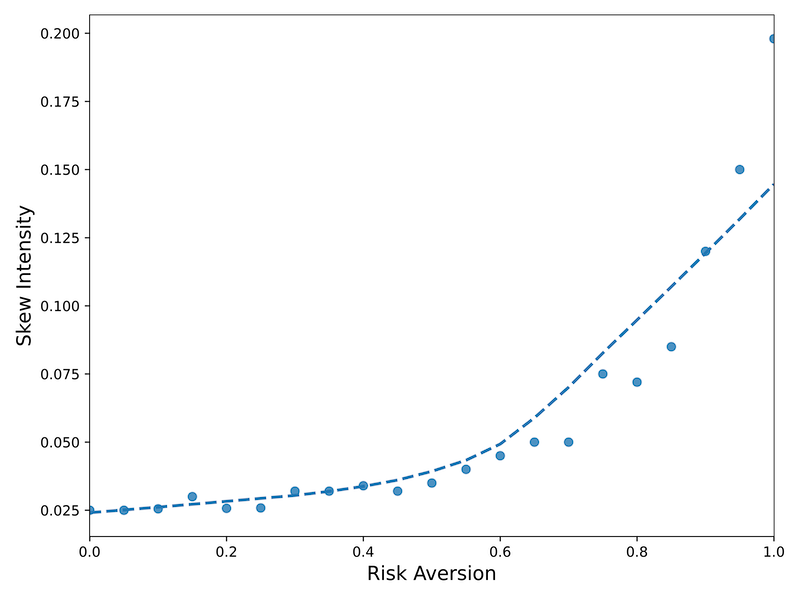}
    \caption{Skew intensity as a function of Risk Aversion}
    \label{fig:Risk Aversion Skew Intensity}
\end{figure}

\noindent \textbf{Setup:} We present in \textit{Figure \ref{fig:Risk Aversion Skew Intensity}}, the skew intensity for different level of risk aversion. The goal of this experiment is to understand if the LP learns to skew more or less its prices based on its type. A more risk averse LP would feel the urgency to exit its position to be less exposed to an unfortunate price move. It will therefore skew mode in order to have higher chances to attract investors and get rid of its inventory.

\noindent \textbf{Analysis:} As we can see in \textit{Figure \ref{fig:Risk Aversion Skew Intensity}}, for a higher level of risk aversion the LP learns to skew more. We observe two regimes, the first one ranging from $0.0$ to $0.6$ where the skew intensity grows linearly as a function of the risk aversion. After, the critical point of $0.6$, we observe a more intense effect of the risk aversion in the skewing of the LP. \textbf{The shared policy of the LPs has then successfully learnt to adapt its pricing strategy as a function of its risk aversion parameter.} By strategically updating its prices, the LP can manage its inventory efficiently in order to maximize its PnL without being exposed to the price risk.

\section{Conclusions and Future Work}

With our new unified RL formulation for both the LP and the LT, via a parametrized family of reward functions expressing the trade-off between PnL and a targeted quantity (specific to each family). The RL-framework we propose and the use of shared policies enable the generalization and the interpolation over a wide range of behaviors for each group of agent. We were able to study the behavior of the different participants in a market simulator composed of a variety of financial agent types providing a more representative setting of the real life market.

The simultaneous learning of the shared policies for two competing groups of RL-based agents raises some interesting questions from a game-theoretic perspective and could be the object of more in-depth future research.

\section*{Disclaimer}
This paper was prepared for informational purposes by the Artificial Intelligence
Research group of JPMorgan Chase \& Co and its affiliates (“J.P. Morgan”), and is not a product
of the Research Department of J.P. Morgan. J.P. Morgan makes no representation and warranty
whatsoever and disclaims all liability, for the completeness, accuracy or reliability of the information
contained herein. This document is not intended as investment research or investment advice, or a
recommendation, offer or solicitation for the purchase or sale of any security, financial instrument,
financial product or service, or to be used in any way for evaluating the merits of participating in
any transaction, and shall not constitute a solicitation under any jurisdiction or to any person, if such
solicitation under such jurisdiction or to such person would be unlawful.
© 2021 JPMorgan Chase \& Co. All rights reserved.

%
\bibliographystyle{ACM-Reference-Format}
\bibliography{investor}


\end{document}